\documentclass[aps,pra,reprint,groupedaddress]{revtex4-1}
\usepackage[latin1]{inputenc}
\usepackage{amsmath}
\usepackage{amsfonts}
\usepackage{amssymb}
\usepackage{graphicx}

\begin{document}
\title{Study of Coherent Perfect Absorption Using Gaussian Beam in a Composite Medium}


\author{Sanjeeb Dey}
\email{snjbde.1@gmail.com}
\affiliation{School of Physics, University of Hyderabad, Hyderabad-500046, India.\\\textbf{snjbde.1@gmail.com} }



\begin{abstract}
There is a study of coherent perfect absorption (CPA) in composite medium (CM) which is made of metal-dielectric composition. The CM, an orthorhombic shaped slab, is illuminated by two identical Gaussian beams (GB) incident at equal angle of incidence and focused on the two interfaces between wall of slab and air from opposite side. It has determine the necessary conditions required for CPA and shows that perfect CPA is possible with GBs. Author also shows broader beams exhibit more absorption than comparatively tightly focused beams.
\end{abstract}
\pacs{}

\maketitle

\section{Introduction}
In recent years there is considerable interest in coherent perfect absorption (CPA) \cite{one,two,three,four,five,six,seven}, super scattering (SS) and critical coupling (CC) \cite{eight,nine,ten,eleven} due to its potential applications in different key areas. 
A closer look at CPA reveals its origin as complete destructive interference and critically coupled, when both the intersecting waves have the same amplitude, while their phases differ by a value of $ \pi $ or it's odd multiple i.e. $\Delta\Phi =(2n+1)\pi$. The higher orders are considering for multiple reflection and transmission in multi-layered system. Such features were discussed \cite{five, six} in recent publications and here we ware discuss some more fundamentals which has various possibilities for a metal-dielectric homogeneous composition \cite{sixteen, seventeen} slab were investigated as composite medium (CM). It was shown that different possible solutions exist even in presence of dispersion with a minimal level of losses (essential for having CPA). However most of the studies on CPA (theories and experiments) till date address only plane waves, while the usual laser sources in the laboratory give out Gaussian beams as output. It is thus necessary to investigate whether CPA like effects persist for such Gaussian input beams. In this paper I address these issues. Replacing the plane wave by a Gaussian beam brings the several complications associated with the angular spectrum decomposition of the beam. For a broad beam, it is shown that both the incident beams can be absorbed almost completely leading to near-null reflection and transmission (or near null scattering). This is quite expected since a broad beam is fairly close to the plane waves. For a tightly focused beam, however, the angular divergence of the propagating k-vector, which are the inheritance plane wave components of the beam increases. As a consequence the reflected and transmitted beams can be highly distorted and the destructive interference for the spatial components is far from perfect. These results are infinite scattering defeating the spirit of CPA.
\begin{figure}[h]
\begin{center}
\includegraphics[width=8.0 cm,height=6.0 cm]{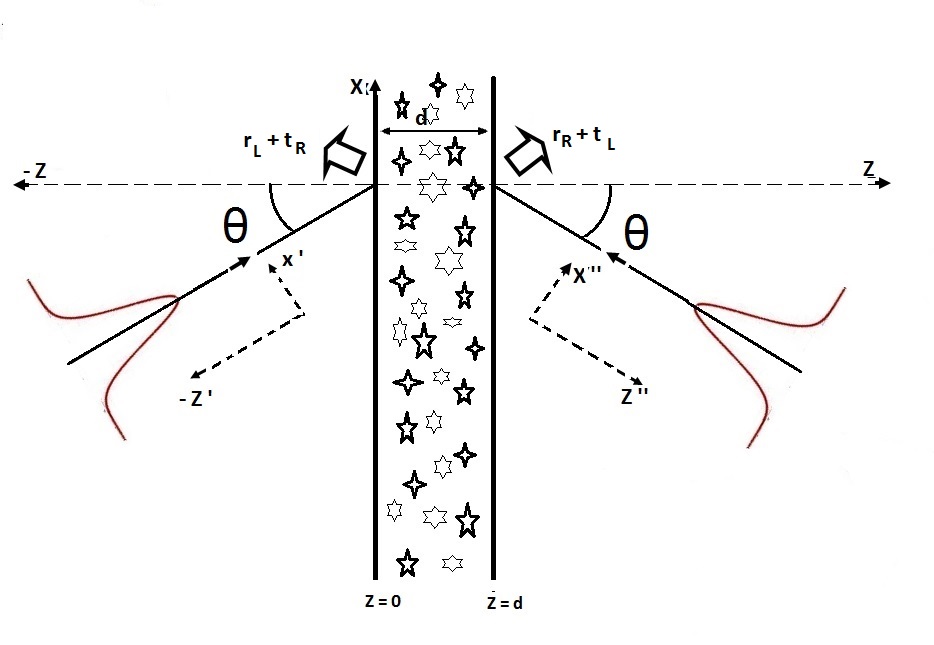}
\caption{Two identical Gaussian beams focus on interface of wall of CM and air. Here, $r_L$, $t_L$, $r_R$ and $ t_R $ are reflection and transmission coefficient of left and right side respectively. The direction of un-prime, prime \& double-prime coordinate system as shown in the figure. }
\end{center}
\end{figure}
The organization of the paper is as follows. In first section, there are description of the system and the illumination geometry. Here, I include the outline of all necessary steps for the angular spectrum decomposition of the Gaussian beams. Assuming a certain polarization state (i.e. s-polarization) and calculating the reflected and transmitted spectra for both the incoming beams. And inverse Fourier transforms lead to the output profiles for the scattered light. Numerical results and discussion are presented in next section. And the last section contains a summary of main results and conclusions.

\section{Formulation \& illumination geometry}
It is considered, two identical Gaussian beams, incident from opposite sides of CM slab at the same incidence angle $ \theta $
 with the z axis (oriented along the normal to the flat surfaces of a CM or CPA medium) (Fig. 1) and $z'$ \& $z{''}$ axis.
 The CPA medium consists of a metal-dielectric composite layer of thickness $``d "$ and is assumed non-magnetic.  The $x-z$ plane at $z = 0$ and $z = d$ are the plane of incidences which are the two interfaces of the CPA medium.
  Henceforth, author uses the label $L$ and $R$ for left and right propagating wave incident from left and right side respectively, so that the resulting reflected and transmitted waves are denoted as $r_L$, $r_R$, $t_L$ and $t_R$ respectively.
   Thus the total scattering amplitude at the interface $z=0$ is given by $r_L+t_R$. This must be same as the total scattering amplitude at the interface $z=d$ given by $r_R+t_L$; since the medium of incidence and emergence are the same due to symmetry of the structure implies $r_L = r_R$ and $t_L = t_R$. In the next section we have discuss the characteristics of CM.
\begin{figure}[h]
\begin{center}
\includegraphics[width=4.0 cm,height=4.0 cm]{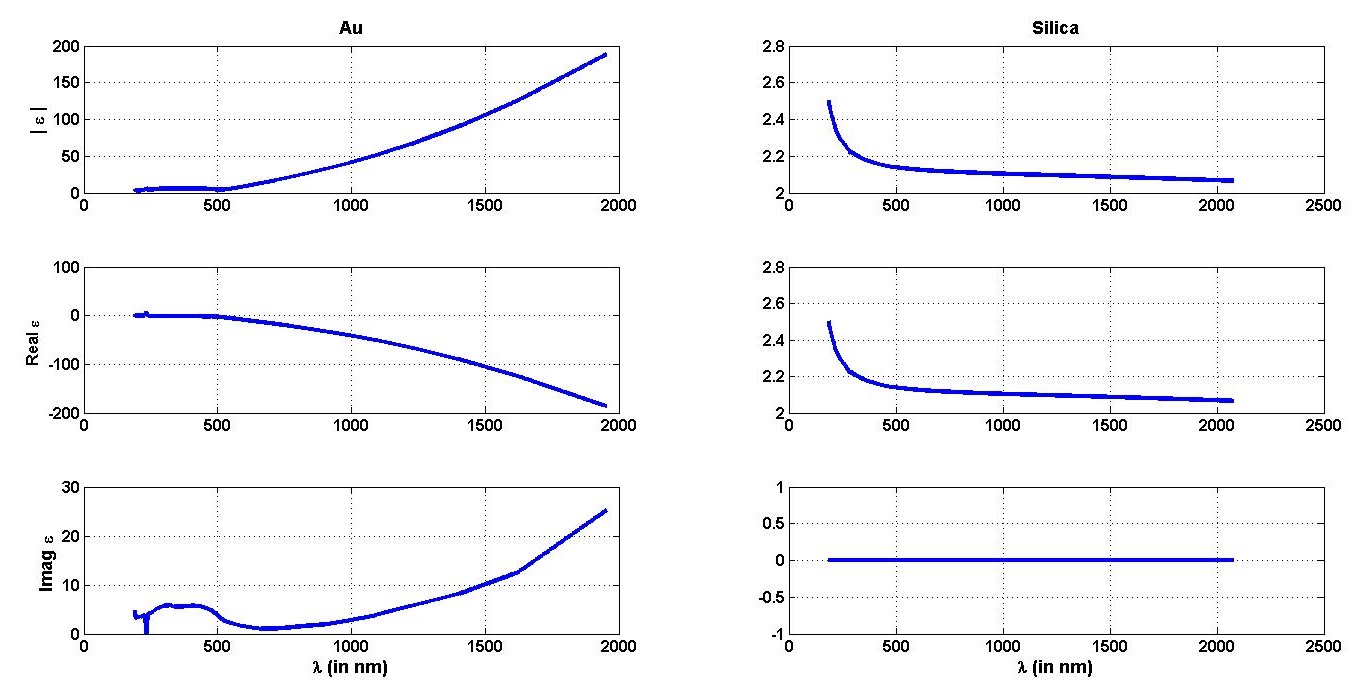}\hspace{0.1cm}\includegraphics[width=4.5 cm,height=4.05 cm]{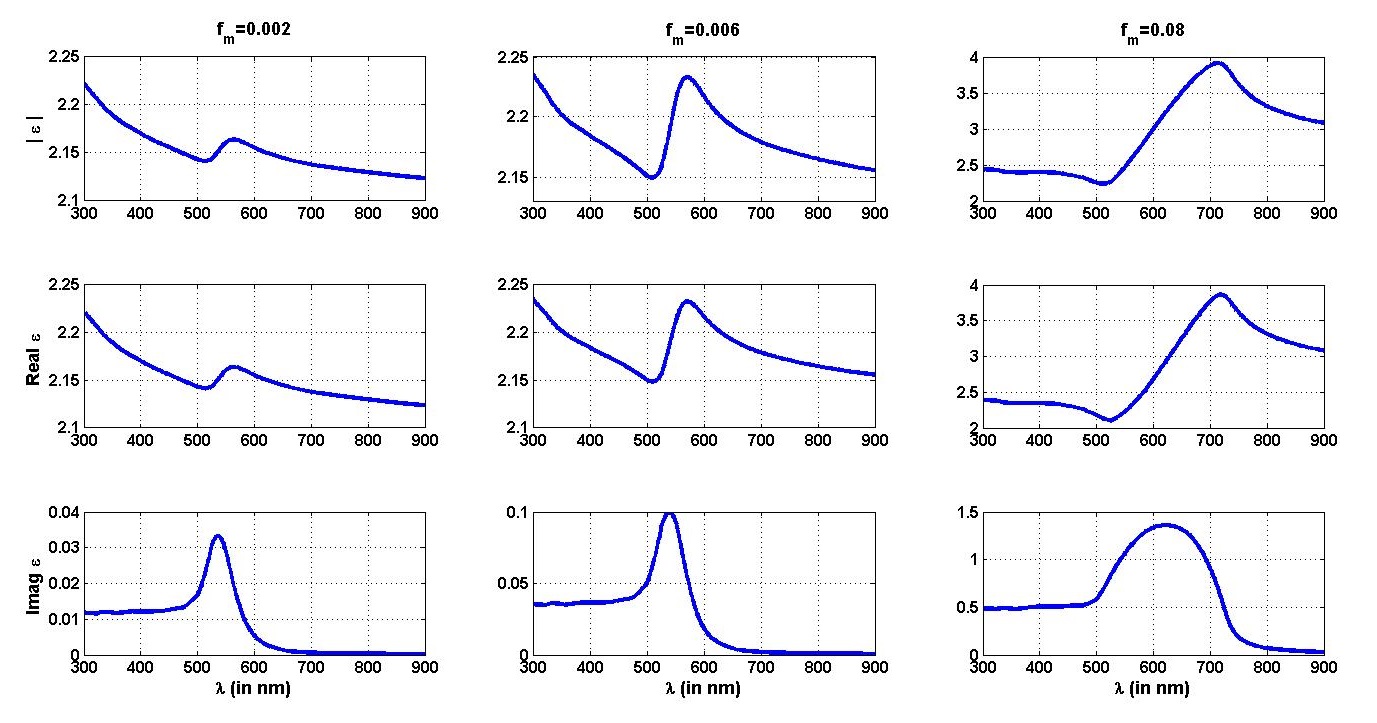}
\caption{First, second and third row of the figure are represent the absolute, real and imaginary values of permittivity of respective medium. First and second column shows gold's ($ Au $) and silica's ($ SiO_2 $) permittivity respectively. Third, fourth and fifth column represent dispersion characteristics of CM in the optical region when volume fractions are ($ f_m= $) 0.002, 0.006 and 0.08 respectively.}
\end{center}
\end{figure}
CM is homogeneous mixture of metal and dielectric. Here, it is made of gold ($ Au $) and silica ($ SiO_2 $). The optical constant of gold \cite{twenty one} and dielectric \cite{twenty} is interpolated \cite{nineteen} and calculated the permittivity for all wavelengths in the optical region. In figure 2 with respect to wavelength the absolute, real and imaginary value of permittivity for metal, dielectric and CM with different volume fraction ($ f_m $) was plotted. To find the effective dielectric function or the permittivity of CM, the Bruggeman effective medium theory (BEMT) \cite{sixteen, seventeen} is being used; which is state as follows.\\$ \epsilon= \frac{1}{4}\lbrace(3f_1-1)\epsilon_1+(3f_2-1)\epsilon_2 $
\begin{eqnarray}
&\pm &\sqrt{(3f_1-1)\epsilon_1+(3f_2-1)\epsilon_2+8\epsilon_1\epsilon_2}\rbrace
\end{eqnarray}
Where, $\epsilon_1$, $f_1$, $\epsilon_2$ and $f_2 = (1-f_1)$ are permittivity and volume fraction of metal and dielectric respectively.
Here, I have shown in the fig 2, characteristic changes in CM in different volume fraction $(f_m=0.002, 0.006$ and $0.08)$ of gold ($ Au $) in the range of 300 to 900 nano-meters, where, the absolute, real and imaginary of permittivity of CM can seen.\\
The CM is illuminated from both side by two identical Gaussian beam sources with equal - intensity, beam spot, wavelength and angle of incident on the CM. For I choose three Cartesian coordinate system, denoted as un-prime, prime and double-prime axis as in the fig 1. Therefore the transformation matrix at $ (x,y,z)=(0,0,0) $  and $ (x,y,z)=(0,0,d) $ are as follows.
\begin{eqnarray}
\begin{bmatrix} x'\\y'\\z' \end{bmatrix} &=& \begin{pmatrix} \cos\theta & 0 & -\sin\theta \\ 0 & 0 & 0 \\ \sin\theta & 0 & \cos\theta \end{pmatrix} \begin{bmatrix} x\\y\\z \end{bmatrix}\\
\begin{bmatrix} x''\\y''\\z'' \end{bmatrix} &=&  \begin{pmatrix} \cos\theta & 0 & \sin\theta \\ 0 & 0 & 0 \\ -\sin\theta & 0 & \cos\theta \end{pmatrix} \begin{bmatrix} x\\y\\z-d \end{bmatrix}
\end{eqnarray}
Then the equations of two such Gaussian beams are propagating along $ z' $ and $ z'' $ axis are represented as follows.
\begin{eqnarray}
E(x'(x,z);z'(x,z)) &=& \frac{w_0}{w(z')}e^{\frac{x'^2}{q'}}e^{ik_0z'}e^{-i\eta(z')}\\ 
E(x''(x,z);z''(x,z)) &=& \frac{w_0}{w(-z'')}e^{\frac{x''^2}{q''}}e^{-ik_0z''}e^{-i\eta(-z'')}
\end{eqnarray}
Where the beam parameters are $ \frac{1}{q'}=\frac{1}{w^2(z')}+\frac{ik_0}{2R(z')} $ and $ \frac{1}{q''}=\frac{1}{w^2(-z')}+\frac{ik_0}{2R(-z')} $. By definition, irrespective of prime or double prime coordinate, the basic beam elements are wave vector, $ k=\frac{2\pi}{\lambda} $; beam waist size,
$ w^2(z)=w_0^2(1+\frac{z^2}{z_0^2}) $; radius of curvature, $ R(z)=z(1+\frac{z_0^2}{z^2}) $; Rayleigh range, $ z_0=\omega_0^2\frac{\pi}{\lambda} $; Gouy phase, $ \eta(z)=\tan^{-1}(\frac{z}{z_0}) $.\\
 
For simplicity, we assume that the amplitude $E$ fields in the beams, have a Gaussian spatial distribution only in the plane of incidence, along the axis perpendicular to the direction of wave propagation and are independent of the coordinate perpendicular to the plane of incidence.
 As there is well known, for a paraxial Gaussian beam profile spreading along $ x' $ \& $ x'' $ axis, 
 the wave front of the beam no longer remain planar but acquire a radius of curvature $R(z')$ and $R(z'')$ at 
 distance $z'$ and $z''$ respectively away from the beam waist (Located at $z' = 0$ and $z''=d$). 
 In other words the Gaussian beam can be regarded as a collection of plane waves of slightly differing (about a central) wave vector. Thus the angular spectrum representation of the Gaussian beam are of the form for $E(x'(x,z);z'(x,z))$  and $E(x''(x,z);z''(x,z))$ 
 \begin{eqnarray}
 \hat{E}(K_{x'};z')&=&\frac{1}{\sqrt{2\pi}}\int_{-\infty}^\infty e^{-iK_{x'}x'}dx'E(x';z')\\
 \hat{E}(K_{x''};z'')&=&\frac{1}{\sqrt{2\pi}}\int_{-\infty}^\infty e^{-iK_{x''}x''}dx''E(x'';z'')
 \end{eqnarray}
After substituting eq (2), eq(4) in eq(6) and eq(3), eq(5) in eq(7) one can reach into following equation.\\$ \hat{E}(K_x;z)= $
\begin{eqnarray}
\frac{w_0}{\sqrt{2\pi}}\int_{-\infty}^\infty dx e^{i(k_0\sin\theta-K_x)x}\frac{e^{\frac{x^2\sin^2\theta}{q}}e^{-i\eta(x\sin\theta)}}{w(x\sin\theta)}
\end{eqnarray}
The incident beams are focused at the interfaces (z = 0 and z = d). The above angular spectrum beam become nothing but infinite number of plane wave which we numerically evaluate. Then each and every such plain waves goes through Fresnel's reflection and transmission coefficient formulas; which gives reflected and transmitted amplitudes at z = 0 and z=d or at the interfaces. Then, added up of one side scattering amplitude (SA) i.e. reflection amplitude with either side transmitted amplitude. At last, inverse Fourier transformation of SA gives the actual beam profile which is comes out which can be written as
\begin{eqnarray}
E(x;z)=\frac{1}{\sqrt{2\pi}}\int_{-\infty}^\infty[SA\hspace{0.08cm} of \hspace{0.08cm}\hat{E}(K_x;z)]e^{iK_xx}dK_x
\end{eqnarray}
The angular spectrum decomposition of the Gaussian beam using Fourier transformation leads to infinite number of plane wave keeping the profile same, because Fourier transformation of a Gaussian is a Gaussian. Each decomposed plane wave obeys Fresnel's reflection and transmission formalism \cite{twenty two}. So each of decomposed beam wave is multiplied by the corresponding reflection and transmission coefficient. Thereafter an inverse Fourier transform leads to the reflected and transmitted beam profile at $z = 0$ and $z=d$ respectively.


\section{Calibrations, Numerical results and discussions}
The CM (nonmagnetic) is taken to be a gold dielectric composite layer of thickness d ($=5 \mu m$). The effective dielectric constant of the composite is calculated using Bruggmann effective theory using the following parameters;
\begin{figure}[h]
\begin{center}
\includegraphics[width=2.8 cm,height=4.0 cm]{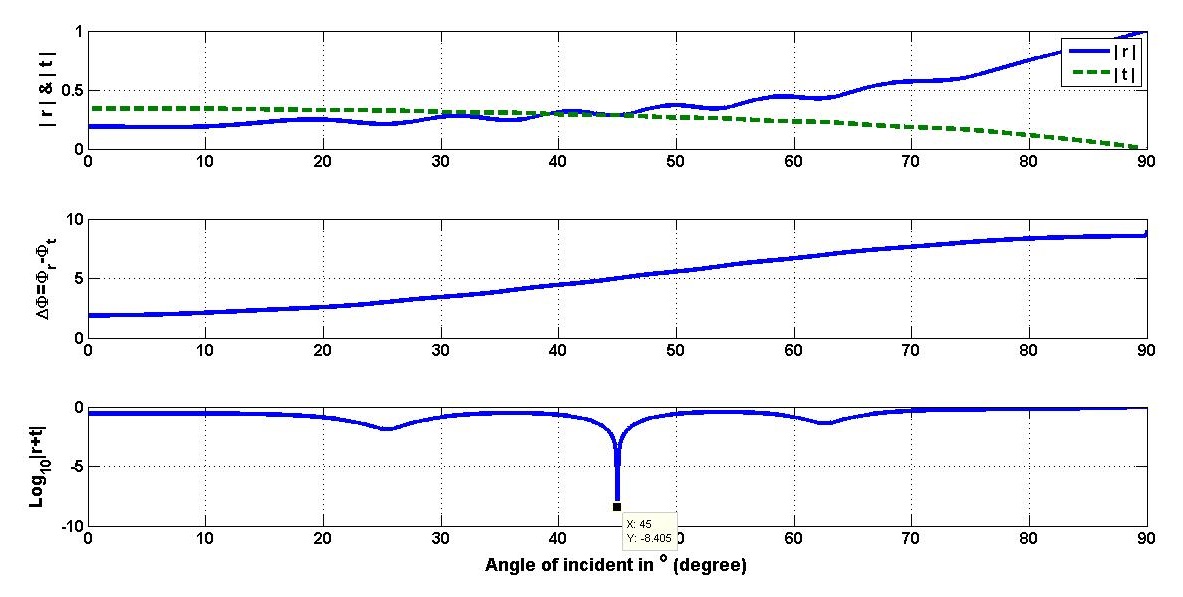}\hspace{0.1cm}\includegraphics[width=2.8 cm,height=4.0 cm]{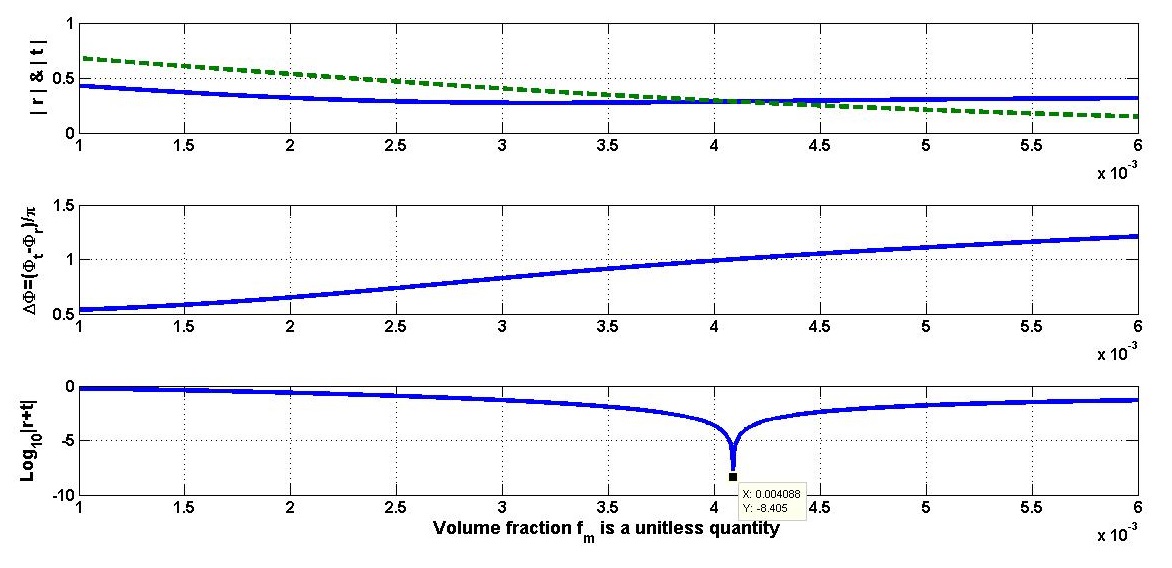}\hspace{0.1cm}\includegraphics[width=2.8 cm,height=4.0 cm]{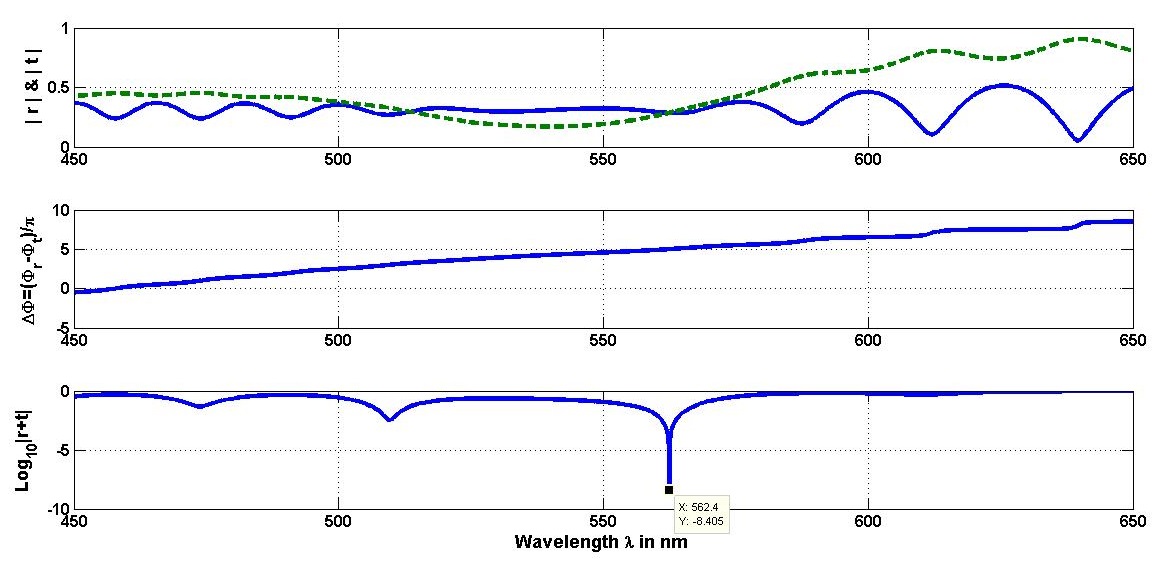}
\caption{From first column of figures states at $ 45^o $ angle of incident, Second column of figures states at $ f_1=0.0040884 $ (unit less) volume fraction and third column of figures states at $ \lambda=562.4 $ nm wavelength  maximum absorption observed with plain wave \cite{five}.}
\end{center}
\end{figure}
a filling factor $f_1=0.004088$ for gold whose dielectric constant $\epsilon_1 (= - 6.4552 + 1.9896 i)$ at a discrete wavelength, $\lambda=562.4$ nm which is taken from Johnson's and Christy's \cite{twenty one} experimental data's interpolating result. The dielectric constant of the dielectric (silica) is, $\epsilon_2 = 2.25$.
 The thickness of the slab $d=5 \mu m$. The calculated permittivity of CM is $\epsilon_{CM} = 2.3225 + 0.0562 i$. An important aspect is to determine of the angle of incidence and the wavelength of the incident fields at which maximum absorption is observed. If one can look into figure 3, I have taken individual observation for \cite{five} angle of incident ($ 45^o $ angle), volume fraction of CM (Gold and Silica) $ f_1=0.0040884 $  and wavelength of two monochromatic infinitely stretched plane wave  $ \lambda=562.4 $ nm fall into CM from opposite side. It is observed, that about $ 10^{-8} $ times scattering amplitude came out comparing to the incident amplitudes. While taking simultaneous observation, the remaining parameters are taking constant.  Now, in the next para, replacing the plane waves with Gaussian beams as state follows. \\
 
 Two identical Gaussian beam both having minimum beam spot size $w_0= 200$ $\mu m$, 
 wavelength $\lambda=562.4$ nm focused on the wall of the slab of CM, is propagating along $z'$ and $-z''$ 
 axis have an incident angle $45^o$  and $ -45^o $ with $z$ axis respectively. In the numerical calculations $K_x=\frac{2\pi}{\Delta x}$ and $ \Delta x=4\frac{w_0}{n} $, where $n=2^{13}$  is taken.
 \begin{figure}[h]
\begin{center}
\includegraphics[width=9 cm,height=6.0 cm]{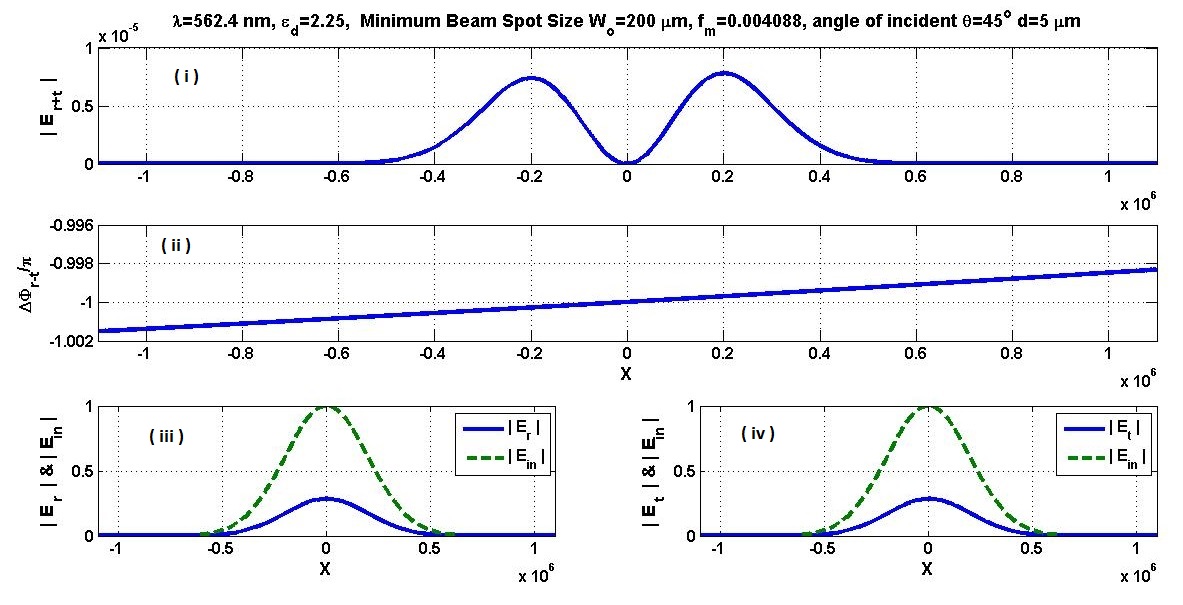}
\caption{(i)Output or resultant beam profile when two identical Gaussian beam focused on the slab from opposite side at $ 45^o $ angle of incident and 562.4 nm wavelength beams. (ii) Phase change along the output beam profile. (iii)Incident and reflected beam profile when the beam is sinning from one side. (iv)Incident and transmitted beam profile when the beam is sinning from one side }
\end{center}
\end{figure}
In the figure 4(i) is the output or resultant beam profile, which can also called scattering amplitude when two identical Gaussian beams ($ \lambda= 562.4$ nm, $ \epsilon_d= 2.25$, $ w_0= 200$ $ \mu $m, $ d=5$ $\mu $m and $ f_1=0.0040884 $) focused from the opposite side on CM. One can see the scattering amplitude is of the order of $10^{-5}$ comparing to the incident amplitude. The next figure 4(ii) shows about $ \pi $ phase change through out the scattering profile amplitude. In figure 4 (iii) and (iv) are showed reflection and transmission beam profile respectively with blue line there are also comparing with incident beam profile is shown in green dotted line while shining the Gaussian beam on CM from one side.\\

In fig 5, it has observed, simultaneously, the output beam profile as well as it's corresponding phase while changing different physical parameters such as width of CM ($ d $), volume fraction ($ f_1 $), wavelength of incident beam ($ \lambda $) and beam waist size ($ w_0 $) respectively when keeping remaining parameters are in constant. 

\begin{figure}[h]
\begin{center}
\includegraphics[width=4.2 cm,height=3.0 cm]{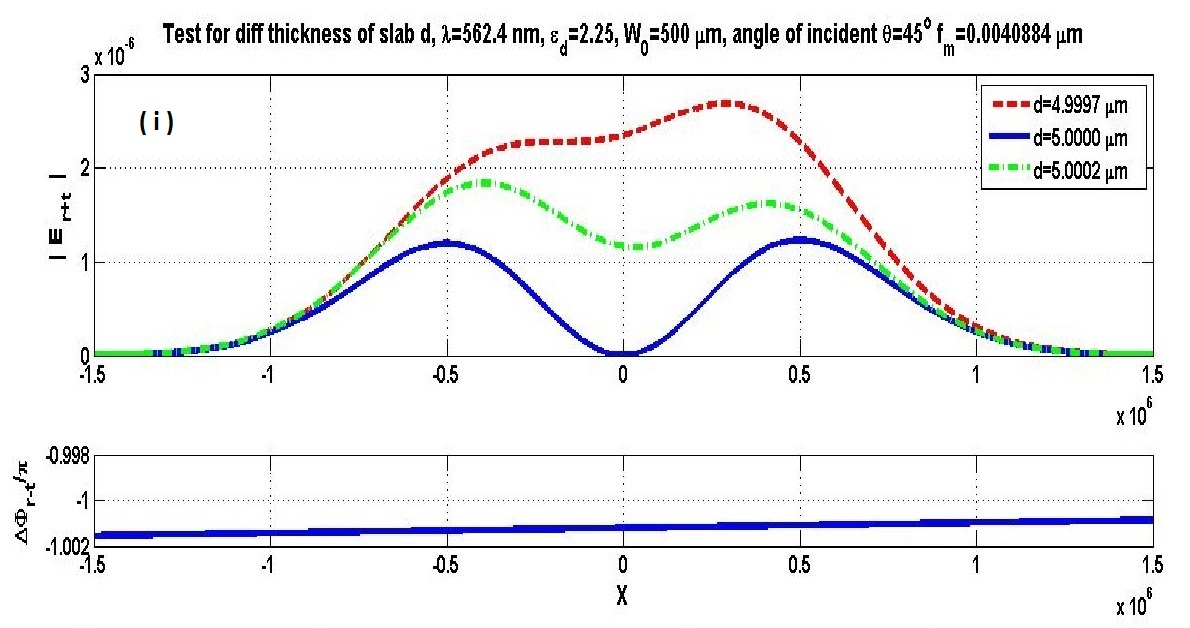}\hspace{0.1cm}\includegraphics[width=4.2 cm,height=3.0 cm]{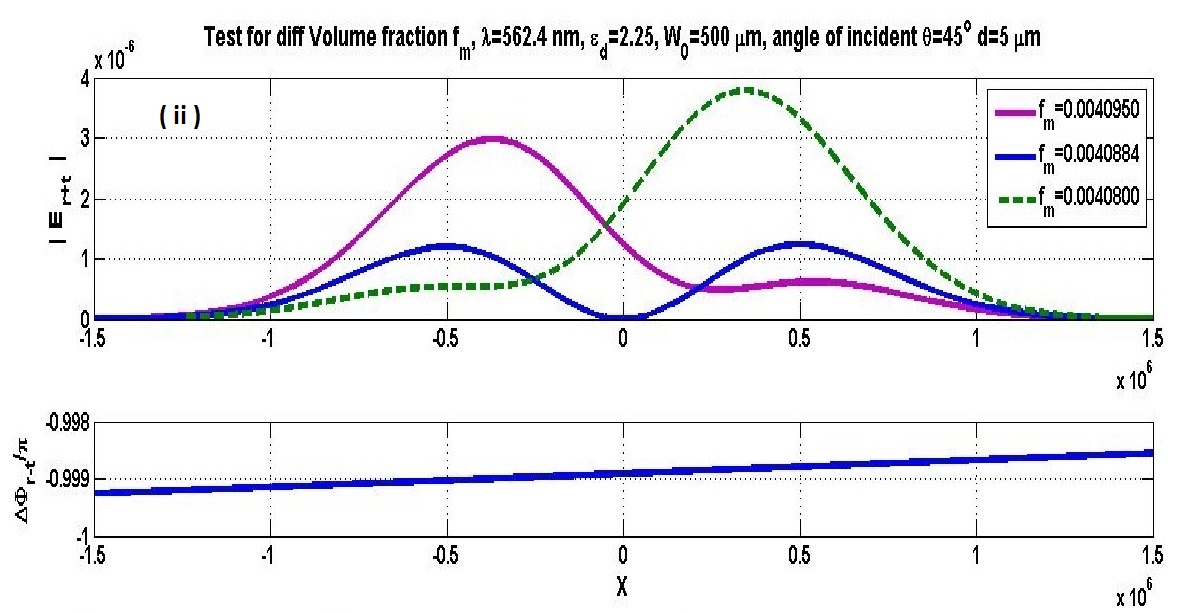}\\ \includegraphics[width=4.2 cm,height=3.0 cm]{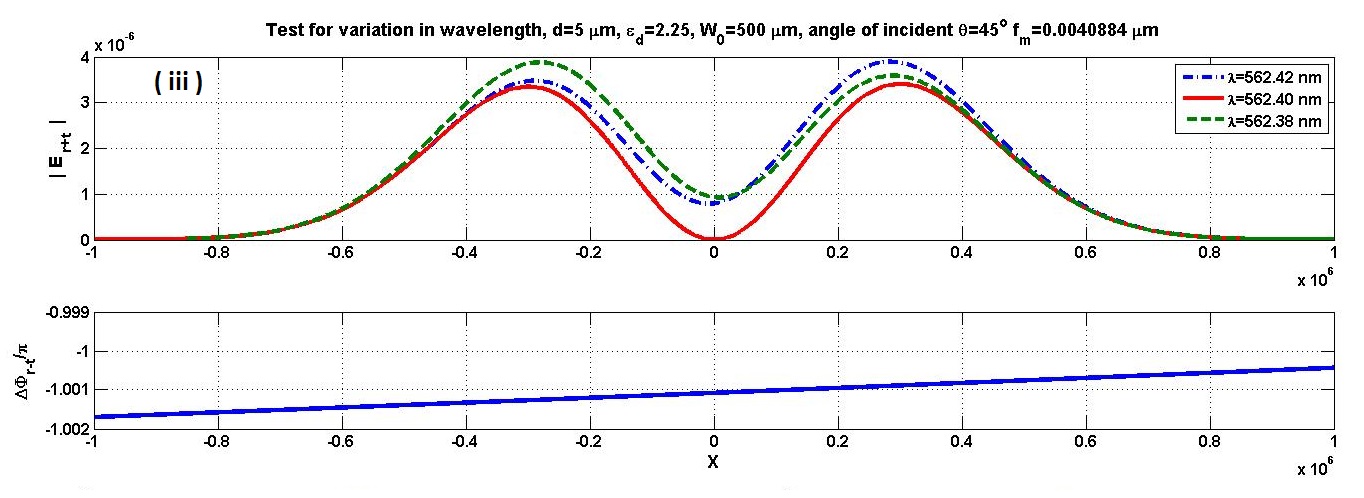}\hspace{0.1cm}\includegraphics[width=4.2 cm,height=3 cm]{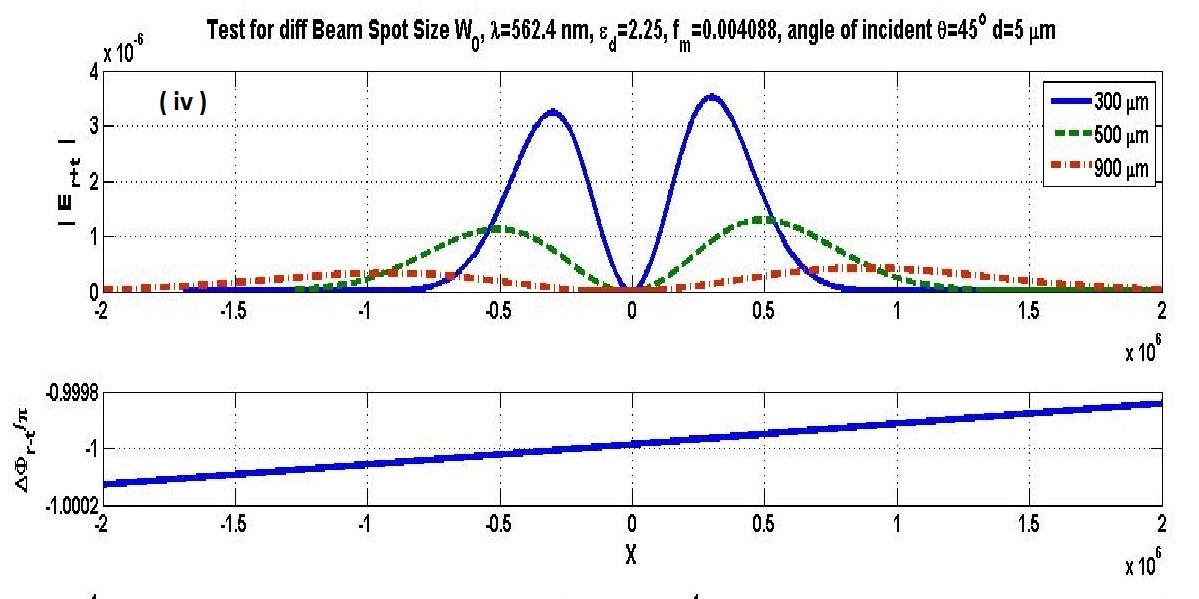}
\caption{ Observation of profile while making (i) small changes of thickness of slab $d=  $4.9997, 5.0000 and 5.0002 $ \mu $m (ii) small variation in volume fraction $ f_m=$0.0040950, 0.0040884 and 0.0040800 , (iii) small variation in wavelength ($ \lambda=$562.42, 562.40 and 562.38  in nm) and (iv) variation in beam waist size ($ w_0=$300, 500 and 900  $ \mu $m)}
\end{center}
\end{figure}
In the fig 5(i), $ d=4.9997 $, $ 5.0000 $ and $ 5.0002 $ in $ \mu $m are taken when $\lambda = 562.4$ nm, $ \epsilon=2.25 $, $ w_0=500 $ $ \mu $m and $ f_1=0.0040884 $ keeps as constant.
We have observed by small changing in volume fraction $ f_1 $ ($= 0.0040950$, $0.0040884$ and $0.0040800$) to observe (fig 5(ii)), the resultant beam front keeping $ d=5 $ $ \mu $m and remaining parameter constant. 
In figure 5(iii), making slight change in wavelength $ \lambda =$ (562.42, 562.40 and 562.38 nm) taken the observation when $ d=5 $ $ \mu $m, $ f_1=0.0040884 $ and remaining parameter constant. And lastly for three different of beam waist size $ w_0 =$ (300, 500  and 900 $ \mu $m) while $ d=5 $ $ \mu $m, $ f_1=0.0040884 $, $ \lambda=562.4 $ nm and the remaining parameters are keeping same.
\section{Conclusion and remarks}
Here, it is studied coherent perfect absorption (CPA) in metal (Gold) - dielectric ($SiO_2$) composite medium (CM) on orthorhombic shaped slab. 
Here, it is considered, two identical Gaussian beams incident on the slab from the opposite end with the same incident angle $ \theta $ with z axis. The permittivity of metal and CM is approximated by Johnson and Christy's ``optical constant of novel metals'' and Bruggeman effective medium theory respectively. I have shown, CPA can be observed with Gaussian beams. Here, the central part and the peripheral parts of the beam is completely absorbed whereas the absorption is reduced in the middle parts of the resultant beam's with incident angle. From the figure 4, we observed a licking of the order of $4\times10^{-5}$ with respect to incident beam amplitude; i.e. minimum 99.9996\% is absorbed of the incident beam due to critical coupling in the composite medium. CPA can occur even for moderately focused Gaussian beams which typically are generated in the laboratories and for utilization in experimental studies of optical problems. A detailed analytic discussion of CPA scattering of fundamental Gaussian beam as a function of different physical parameters of the system was carried out. These results and observations can use in many physics and optical problem.

\textbf{Acknowledgements}\\ Author is thankful to Suneel Singh for his useful suggestions and discussions. He is also thankful to UGC (Govt. of India)  for financial support.
\bibliography{basename of .bib file}

\end{document}